# Analytic Solution of Emden-Fowler Equation and Critical Adsorption in Spherical Geometry

S. Gnutzmann and U. Ritschel

*Fachbereich Physik, Universität GH Essen, 45117 Essen (F R Germany)*

## Abstract

In the framework of mean-field theory the equation for the order-parameter profile in a spherically-symmetric geometry at the bulk critical point reduces to an Emden-Fowler problem. We obtain analytic solutions for the surface universality class of extraordinary transitions in $d = 4$ for a spherical shell, which may serve as a starting point for a pertubative calculation. It is demonstrated that the solution correctly reproduces the Fisher-de Gennes effect in the limit of the parallel-plate geometry.

Typeset using REVTEX

1

Surface critical phenomena have received a great deal of attention during recent years, both theoretically [1] and experimentally [2]. In particular, critical adsorption, which occurs near a critical point when a wall favours one of the phases, has been discussed intensively in the literature [3]. Theoretically the phenomenon is identified with the universality class of extraordinary transitions [1]. One of the most important quantities to be determined in this context is the order-parameter profile in some given geometry subject to specified boundary conditions. In this letter we present *analytic* solutions to the mean-field (MF) equations for the order parameter at the bulk critical point in spherically-symmetric geometry and for dimension $d = 4$.

From the *physical* point of view, knowledge of MF solutions, especially for $d = 4$, is desirable for several reasons. Firstly, the MF theory is exact in $d \geq 4$ and, thus, it provides the presumably simplest model for critical phenomena in non-trivial geometries. Secondly, the MF solutions are the basis for more sophisticated techniques like the $\epsilon$-expansion by which, for example, Diehl and Smock [4] obtained profiles for $d = 3$ in the semi-infinite geometry. Thirdly, and this has been the main rationale for our present efforts, analytic solutions may serve as the starting point for other geometries by exploiting conformal invariance at the critical point. Some time ago, Burkhardt and Eisenriegler [5] calculated in this way the order parameter in the interior of a single sphere at bulk criticality, starting from the well-known result $m \sim z^{-x_\phi}$ of the semi-infinite case, where $x_\phi$ is the scaling dimension of $m$. Now, the geometry of two concentric spheres is conformally equivalent to two separate spheres; the interior of the original configuration can be mapped conformally to the exterior of the two-sphere geometry. Thus, by solving the former highly symmetric problem, one has simultaneously treated the other more complicated configuration, which, in turn, is of interest in the context of Casimir forces between spherical objects immersed in a critical fluid [6].

But also from the *mathematical* point of view the MF equations in spherical geometry are interesting and non-trivial. Taking into account the radial symmetry, the order-parameter field at the critical point has to satisfy

$$\ddot{m} + \frac{d-1}{r} \dot{m} - m^3 = 0 \,. \tag{1}$$

This equation is a special case of the generalized Emden-Fowler equation

$$\ddot{m} + \beta(r)\,\dot{m} + \alpha(r)\,m^k = 0 \qquad k \neq -1 \tag{2}$$

as discussed by Sarlet and Bahar [7]. Nonlinear second-order equations of the Emden-Fowler type have a long history in the literature [8]. Originally, they were introduced in astrophysics to describe gas spheres subject to the laws of thermodynamics and their own gravitational forces.

Sarlet and Bahar found [7] that a first integral to (2) exists if the functions $\alpha$ and $\beta$ satisfy the constraint equation

$$\alpha^{-2/(k+3)}\exp\left(-\frac{4}{k+3}\int^r \beta(r')\,dr'\right) - A\int^r \exp\left(-\int^{r'}\beta(r'')\,dr''\right)dr' = B\,, \tag{3}$$

where $A$ and $B$ are constants. If now the concrete form (1) is taken, one finds that the constraint (3) for $k = 3$ is satisfied only when $d = 4$.

Concentrating in the following on $d = 4$ and $k = 3$ the first integral to (2) is given by

$$I = 2\,r^4\,\dot{m}^2 + 4\,r^3\,m\,\dot{m} - r^4\,m^4\,. \tag{4}$$

As said above, we are seeking profiles relevant for the extraordinary transition. Thus, the order parameter must be enhanced close to the walls and should have one minimum in between. Employing this minimum condition, we find

$$m_{\min}^4\,r_{\min}^4 = -I\,, \tag{5}$$

and, as the left-hand side is always positive, it follows for finite $m_{\min}$ and $r_{\min}$ that $I < 0$. The case $I = 0$ is best be treated separately. As we shall see below, it leads to profiles for a single spherical surface.

In order to determine solutions of (4), we use the substitution $m = u/r$ and obtain a separable first-order equation for $u$. From the integral of this equation, $m$ and the radial coordinate $r$ can be determined in the parametric form

$$m(\varphi) = \frac{\sqrt{\gamma - 1}}{r(\varphi)\,\cos\varphi} \qquad \text{and} \qquad r(\varphi) = R_0 \exp\left(\frac{F(\varphi,k)}{\sqrt{\gamma}}\right) \tag{6}$$

with constants $\gamma = \sqrt{1 - I}$ and $k = \sqrt{(\gamma + 1)/2\gamma}$. The function $F$ in (6) is an elliptic integral of the first kind, defined by

$$F(\varphi, k) = \int_0^\varphi \frac{d\varphi'}{\sqrt{1 - k^2 \sin^2 \varphi'}}\,.$$

The parameter $\varphi$ varies in the interval $[-\pi/2, \pi/2]$. It is straightforward to show that $m$ becomes singular at the end points $\pm\pi/2$; the corresponding radii are

$$R_\pm = R_0 \exp\left(\pm \frac{F(\pi/2, k)}{\sqrt{\gamma}}\right). \tag{7}$$

Further, the distance between singularities is

$$D \equiv R_+ - R_- = 2 R_0 \sinh\left(\frac{F(\pi/2, k)}{\sqrt{\gamma}}\right), \tag{8}$$

and the constant $R_0$ turns out to be the geometric mean of $R_+$ and $R_-$. For fixed $R_0$, the distance $D$ is monotonously decreasing as a function of $\gamma$. $D \to \infty$ for $\gamma \to 1$, and $D \to 0$ for $\gamma \to \infty$. The minimal value is always located at some $r_{\min} > R_0$, and the corresponding parameter value $\varphi_{\min}$ is determined by

$$\cos \varphi_{\min} = \left(\frac{\gamma - 1}{\gamma + 1}\right)^{1/4}.$$

A number of solutions for fixed $R_0 = 1$ and various values of $\gamma$ is depicted in our figure.

In the extreme case of infinite surface enhancement, corresponding to the fixed point in the renormalization-group calculation [1], the radii $R_-$ and $R_+$ may be identified with inner and outer surface, respectively. This is also the situation, where conformal invariance can be directly employed to transform to other geometries [5]. In the more general case of a (spherical) surface at $r = R_s$ and *finite* enhancement at the surface, the solution has to satisfy boundary conditions of the form

$$\partial_r m\big|_{r=R_s} = c\, m\big|_{r=R_s}. \tag{9}$$

If we have $R_- < R_s < R_0$, then, expressed in terms of the parameter $\varphi$, the constant $c$ can be written as

$$c = \tan \varphi_s - \left(\frac{2}{2\gamma + (\gamma + 1)\sin^2 \varphi_s}\right)^{1/2} \tag{10}$$

with $r(\varphi_s) = R_s$. For $\varphi_s \to -\pi/2$ the constant $c$ tends to $-\infty$.

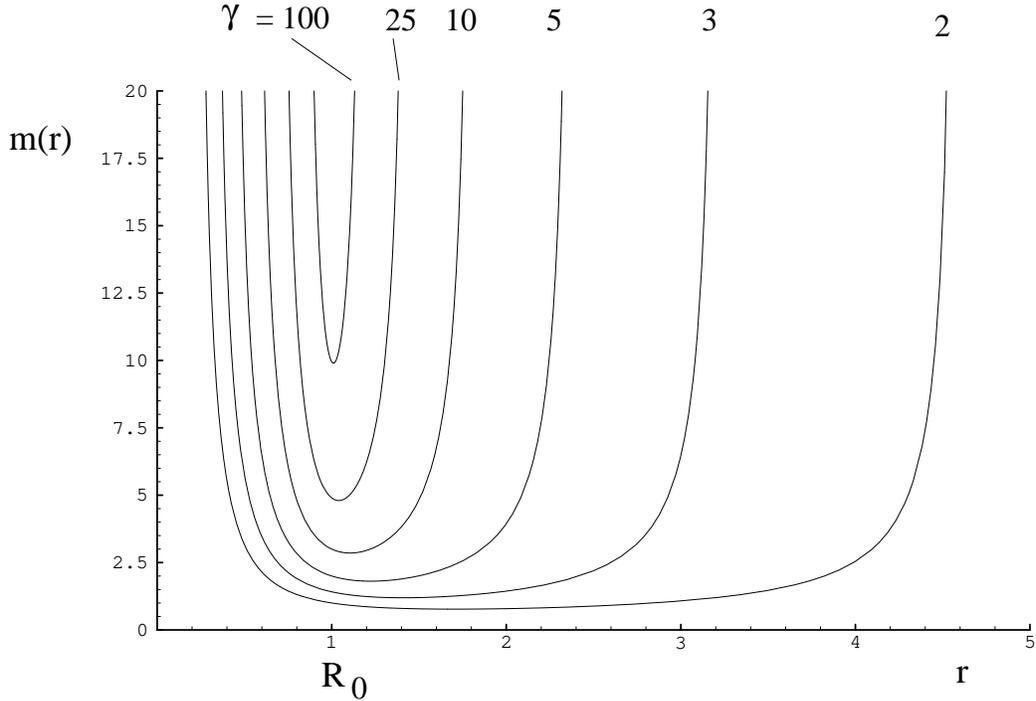

**Fig.:** Profiles from solution (6) with $R_0 = 1$ and various values of $\gamma$.

An important question, which has received considerable attention in the literature [9], is how the form of profiles close to one surface is influenced by a second more distant surface. Fisher and de Gennes [10] conjectured that in the parallel-plate geometry the profile has the expansion

$$m \sim z^{-x_\phi} \left\{ 1 + \text{const} \left( \frac{z}{D} \right)^d + \text{higher orders} \right\}, \tag{11}$$

where $z$ is the perpendicular distance to the surface and $x_\phi$ is the scaling dimension of $m$. In the MF case we have $x_\phi = 1$. Further, Cardy [11] determined the leading correction to the profile due to curvature of the surface. He found

$$m \sim z^{-x_\phi} \left\{ 1 - \frac{x_\phi z}{4} \left( \frac{1}{R_1} + \frac{1}{R_2} \right) + \text{higher orders} \right\}, \tag{12}$$

where $R_1$ and $R_2$ are the principle radii of curvature. The analogous expression for the special case of a sphere can be also derived by expanding the exact profiles of Burkhardt and Eisenriegler [5].

Now, our solution (6) may be compared with both (more general) results mentioned above. We restrict ourself to the case of infinite surface enhancement $c \to -\infty$. Close to $R_-$, the order parameter has the expansion

$$m = \frac{\sqrt{2}}{r - R_-} \left\{ 1 - \frac{\epsilon}{2} + \frac{\epsilon^2}{4} - \frac{\epsilon^3}{8} + \left(1 + \frac{2\gamma^2 - 2}{5}\right) \frac{\epsilon^4}{16} + \mathcal{O}\left(\epsilon^5\right) \right\} \tag{13}$$

with $\epsilon \equiv (r - R_-)/R_-$. In this expansion all terms up to order $\epsilon^3$ are pure curvature terms, i.e., they are not due to the second surface at $R_+$. In particular, the first-order term is consistent with Cardy's result (12). In the limit $R_- \to \infty$ and $D$ fixed, the limit of parallel plates, (13) reduces to

$$m = \frac{\sqrt{2}}{z} \left\{ 1 + \frac{1}{5} \left[F\left(\pi/2, \sqrt{2}/2\right)\right]^4 \left(\frac{z}{D}\right)^4 + \text{higher orders} \right\} \tag{14}$$

with $z = r - R_-$. This is in agreement with the prediction of Fisher and de Gennes (11). For the parallel-plate geometry and for general $z \in [0, D]$ our solution (6) takes the simple parametric form

$$m(\varphi) = \frac{m_0}{\cos \varphi} \quad \text{and} \quad z(\varphi) = \frac{D}{2} \left(1 + \frac{F(\varphi, \sqrt{2}/2)}{F(\pi/2, \sqrt{2}/2)}\right), \tag{15}$$

where $m_0$, the value in the center between the plates, is given by $m_0 = 2F(\pi/2, \sqrt{2}/2)/D$. A solution equivalent to (15) was derived in [12].

Eventually, for the much simpler special case $\gamma = 1$ (or $I = 0$ in (4)) one can calculate explicit solutions of (1) of the form

$$m = \frac{2\sqrt{2}\,R}{R^2 - r^2} \quad \text{and} \quad m = \frac{2\sqrt{2}\,R}{r^2 - R^2} \tag{16}$$

which describe a critical medium enclosed in a sphere (with radius $R$) and a single sphere (with radius $R$) immersed in a critical medium, respectively. Again, (16) agrees with Burkhardt and Eisenriegler [5].

*Acknowledgements:* We wish to thank H. W. Diehl, E. Eisenriegler, and M. Krech for helpful comments and hints to the literature. Further, we are grateful to the Deutsche Forschungsgemeinschaft for partial support through Sonderforschungsbereich 237.